\documentclass[11pt,a4paper]{iopart}
\usepackage{varioref}
\usepackage[dvips]{graphicx}
\usepackage[latin1]{inputenc}
\usepackage{iopams}

\newcommand{\AmS}{{\protect\the\textfont2
  A\kern-.1667em\lower.5ex\hbox{M}\kern-.125emS}}

\begin{document}

\title[Identified particle production in p+p and d+Au collisions at RHIC]
{Identified particle production in p+p and d+Au collisions at RHIC}

\author{Hongyan Yang (for the BRAHMS Collaboration\footnote[1]
{For a full author list and acknowledgments, see the appendix of this volume.})}

\address{Department of Physics and Technology,
University of Bergen, \\ Allegaten 55, 5007 Bergen, Norway} \ead{hongyan@ift.uib.no}

\begin{abstract}
The BRAHMS experiment at RHIC has measured the transverse momentum
spectra of charged pions, kaons and (anti-)protons over a wide range of
rapidity in d+Au and p+p collisions at $\sqrt{s_{NN}}=200~$GeV. The
nuclear modification factor $R_{dAu}$ at forward rapidities shows a
clear suppression for $\pi^{+}$. The measured net-proton yields in p+p
collisions are compared to PYTHIA and HIJING/B and seem to be better
described by the latter. 
\end{abstract}

%\pacs{25.75.-q, 25.75.Nq}
\submitto{\JPG}

\section{Introduction}
Ultra-relativistic p+A collisions are used to understand the initial
state effects in heavy-ion collisions, which may have produced
deconfined nuclear matter ${~\cite{QGP03}}$. By the measurement of
hadrons produced in relativistic d+Au and p+p collisions over a wide
range of rapidity, one can try to disentangle the modification of the
parton distributions in nuclei (e.g. shadowing) and the change of
$p_T$ spectra of produced particles caused by initial and final state
multiple scattering in cold nuclear matter (e.g. Cronin effect
${~\cite{Cronin}}$) and thus constrain various dynamical evolution
scenarios and initial conditions (e.g. Color Glass Condensate
${~\cite{CGC}}$).

Experimental data from d+Au collisions at RHIC from the BRAHMS, PHENIX
and STAR Collaborations ${~\cite{BRAHMSdApaper, PHENIXdApaper,
STARdApaper}}$ has verified the prediction of the Cronin effect at
mid-rapidity, i.~e. the enhancement of intermediate-$p_T$ hadrons in
d+Au collisions as compared to p+p, which might be due to the initial
state multiple parton scattering. At forward rapidity $\eta
\sim 3.2$ the suppression of hadron spectra is consistent with
predictions by the CGC model ${~\cite{CGCKKT}}$. The Cronin
enhancement at mid-rapidity and the high-$p_T$ suppression at forward
rapidity are characterized by the nuclear modification factor
$R_{dAu}$, which is defined as a ratio of the invariant particle
spectra from d+Au collisions to reference spectra in p+p collisions
scaled by the average number of binary nucleon-nucleon collisions in
d+Au ($<N_{coll}>$): 

\begin{equation}
R_{dAu}=\frac{d^2N_{dAu}/2\pi p_Tdydp_T}{<N_{coll}>\times d^2N_{pp}/2\pi p_Tdydp_T}.
\end{equation}

\section{Analysis}
BRAHMS (Broad Range Hadron Magnetic Spectrometers) has two rotatable
magnetic spectrometers with particle identification capabilities for
charged hadrons, which allow the study of particle production in a
broad range of both transverse momenta and rapidities. 
\begin{figure}[htbp] \centering
 \includegraphics[width=0.75\columnwidth]{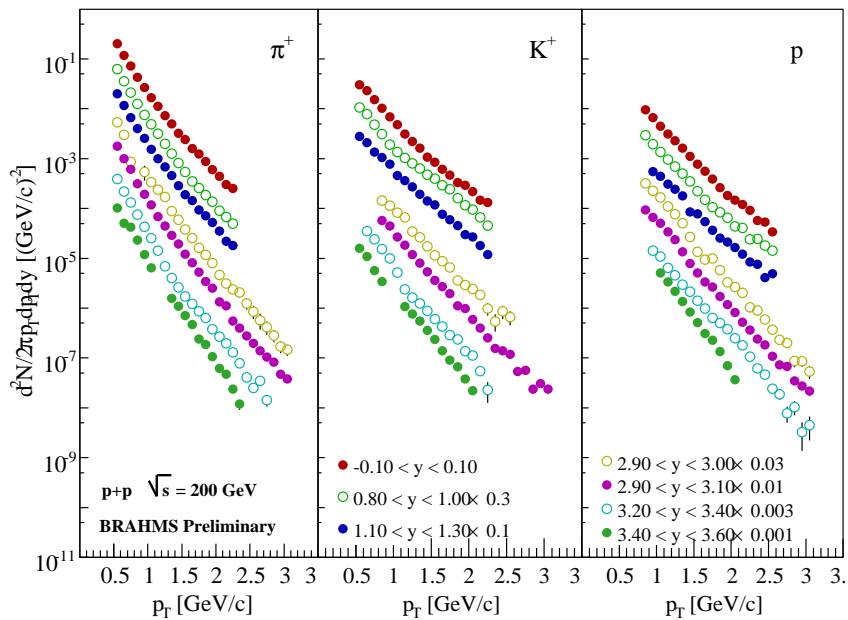}
  \caption{Invariant $p_{T}$ spectra of identified particles at
different rapidities (from 0 to 3.5) in p+p collisions at
$\sqrt{s}=200~$GeV. Only the spectra of positive particles are
shown. Spectra at different rapidities are scaled for clarity.} 
  \label{fig:spectra}
\end{figure}

 BRAHMS uses the Time-of-Flight (TOF) technique in the mid-rapidity
spectrometer (MRS) and the Forward Spectrometer (FS) and a Ring
Imaging \v{C}erenkov (RICH) detector at the back of the FS for the
identification of particles with high momentum. In this analysis pions
and kaons are separated up to 1.8~GeV/c in momentum by the TOF in the
MRS, and the separation of pions and kaons in the FS has been extended
up to 25-30~GeV/c by the RICH. Details on the BRAHMS  detector system can
be found in ~\cite{Brah03, NIM07}.

Spectra in both the MRS and the FS spectrometers have been corrected
for geometric acceptance, tracking and PID efficiencies, in-flight
decay, absorption and multiple scattering. The normalization is done
relative to the number of events in minimum bias data triggered by the
global detectors, and several settings are combined. Weak decays and
feed-down have not been corrected for the protons and anti-protons . 
The resulting invariant spectra in p+p collisions at
different rapidities are shown in Figure~\ref{fig:spectra}. The
invariant transverse momentum spectra in d+Au collisions at different
centralities have already been shown in ~\cite{Hyqm05}. 
  \begin{figure}[htbp] \centering
    \includegraphics[width=0.90\columnwidth]{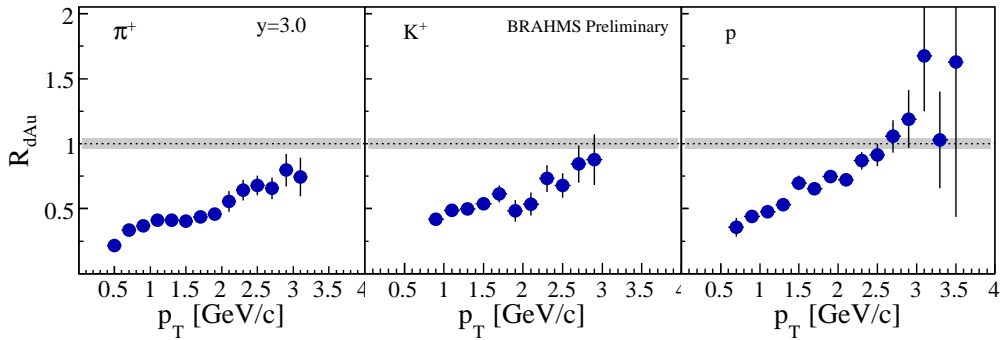}
    \caption{$R_{dAu}$ of $\pi^{+}$, $K^{+}$ and protons at forward
    rapidity $y=3.0$ in minimum bias d+Au collisions ($<N_{coll}>=7.2$).
    A 8\% systematic error is included.}
    \label{fig:rdau}
  \end{figure}

\section{Nuclear modifaction factor}
The nuclear modification factor (Figure~\ref{fig:rdau}) for identified
particles at forward rapidity in minimum bias d+Au collisions was
constructed using $<N_{coll}>=7.2$. 
A clear suppression for $\pi^+$ is observed and $R_{dAu}$ reaches $\sim~0.7$ at $p_T\sim~2.5$~GeV/c.
The suppression of $\pi^+$ is comparable to that for negative hadrons
$h^-$ where negative pions are the main contribution
${~\cite{BRAHMSdApaper}}$. 
Kaons show a similar suppression while protons cross $R_{dAu}=1$ (no
suppression) at $p_T~\sim~2.5~$GeV/c. 

\section{Stopping}

The rapidity dependence of the net-proton yield in d+Au and p+p
collisions may shed light on the stopping and baryon number transport
process in nuclear collisions. The net-proton transverse momentum
spectra are constructed by subtracting the anti-proton spectra from the
proton spectra $p_T$-bin-by-$p_T$-bin. Due to our limited acceptance
at low transverse momenta (see Fig.~\ref{fig:spectra}), the yields at
low $p_T$ have to be determined by fitting to the data a fit-function
and extrapolating the function to low $p_T$. Within our acceptance
both a Boltzmann function and an exponential in $p_T$ describe the
data equally well. Therefore, both function were used for the
extrapolation which results in a lower (Boltzmann) and upper
(exponential in $p_T$) limit for the integrated yield. The functions
were extrapolated to the low $p_T$ region to calculate the integrated
yield at low $p_T$, which was then combined with the yield calculated
from data by summing up the $p_T$-bins in order to obtain the total
net-proton rapidity density over the full $p_T$ range.

\begin{figure}[htbp] \centering
    \includegraphics[width=0.58\columnwidth,angle=0]{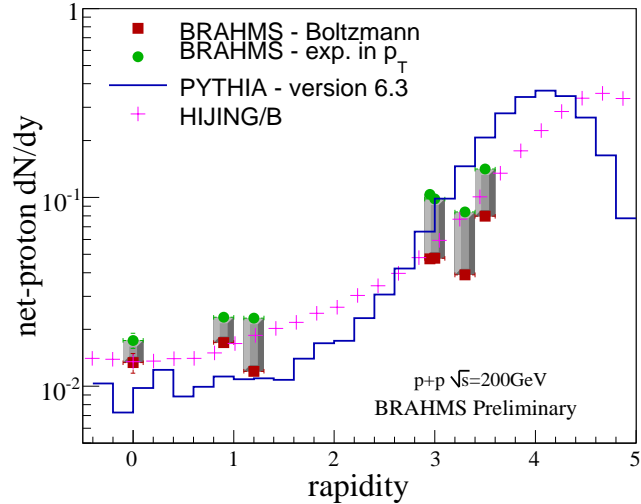}
    \caption{Net-proton rapidity density in p+p
collisions. The squares are the net-proton yields obtained by using a Boltzmann
function, and the dots are those obtained using an exponential
function in $p_T$. The gray boxes between the two sets of
extrapolations indicate the range of solutions which cannot be
distinguished by our data.} 
    \label{fig:netproton}
  \end{figure}

Figure~\ref{fig:netproton} shows the net-proton rapidity distribution
in p+p collisons at 200~GeV and a comparison to PYTHIA
${~\cite{PYTHIA}}$ and the HIJING model with baryon junction
${~\cite{HIJINGB}}$ including single diffractive processes. The
squares are the results of a Boltzmann extrapolation, while the dots
represent those from an exponential function in $p_T$.

The region around mid-rapidity (between 0 and 1) is almost
baryon-free, while a large net-proton density is observed at forward
rapidity (around 3). Even though we have an uncertainty of about 50\%
due to the extrapolation procedure  
HIJING/B's estimate is closer to our data, while PYTHIA's is
systematically lower at mid-rapidity and higher at forward.

\end{document}